\documentclass[12pt,epsfig]{amsart}
\usepackage{amsbsy,amssymb,amscd,amsfonts,latexsym,amstext,delarray,
amsmath,epsfig,graphicx}  \headsep=15pt
\usepackage[ansinew]{inputenc}
\usepackage{hyperref}

\newtheorem{thm}{Theorem}[section]
\newtheorem{prop}[thm]{Proposition}

\newtheorem{lem}[thm]{Lemma}

\newtheorem{rem}[thm]{Remark}

\numberwithin{equation}{section}

\def\C{{\mathbb C}}

\renewcommand{\H}{{\mathbb H}}
\def\N{{\mathbb N}}

\def\Z{{\mathbb Z}}
\def\R{{\mathbb R}}

\def\cA{{\mathcal A}}
\def\cB{{\mathcal B}}
\def\cC{{\mathcal C}}

\def\cE{{\mathcal E}}

\def\cH{{\mathcal H}}

\def\cL{{\mathcal L}}

\newcommand{\ie}{{\it i.e.\/}\ }

\newcommand{\cf}{{\it cf.\/}\ }

\def\text{\hbox}

\def\Aut{{\rm Aut}}

\def\End{{\rm End}}

\def\Hom{{\rm Hom}}

\def\Int{{\rm Int}}

\def\SU{{\rm SU}}

\def\Tr{{\rm Tr}}

\def\qqq{\,,\quad \forall\,}

\title
{Why the Standard Model}

\author[Chamseddine]{Ali H. Chamseddine}
\author[Connes]{Alain Connes}

\address{ A.~Chamseddine: Physics
Department, American University of Beirut, Lebanon\\
and I.H.E.S.} \email{chams@aub.edu.lb}
\address{A.~Connes: Coll\`ege de France \\
3, rue d'Ulm \\
Paris, F-75005 France\\
I.H.E.S. and Vanderbilt University}
 \email{alain@connes.org}

\begin{document}

\begin{abstract} The Standard Model is based on the gauge invariance principle with
gauge group ${\rm U}(1)\times \SU(2)\times \SU(3)$ and suitable
representations for fermions and bosons, which are begging for a
conceptual understanding. We propose a purely gravitational
explanation: space-time has a fine structure given as a product of a
four dimensional continuum by a finite noncommutative geometry F.
The raison d'être for F is to correct the K-theoretic dimension from
four to ten (modulo eight). We classify the irreducible finite
noncommutative geometries of K-theoretic dimension six and show that
the dimension (per generation) is a square of an integer k. Under an
additional hypothesis of quaternion linearity, the geometry which
reproduces the Standard Model is singled out (and one gets $k=4$)
with the correct quantum numbers for all fields. The spectral action
applied to the product $M\times F$ delivers the full Standard Model,
with neutrino mixing,  coupled to gravity, and makes predictions
(the number of generations is still an input).
\end{abstract}

\maketitle

\section{Introduction}

The Standard Model is based on the gauge invariance principle with
gauge group $$G={\rm U}(1)\times \SU(2)\times \SU(3)$$ and suitable
representations for fermions and bosons. It involves additional
scalar fields, the Higgs fields and a number of key mechanisms such
as V$-$A, spontaneous symmetry breaking etc... While the values of
the hypercharges can be inferred from the condition of cancelation
of anomalies, there is no conceptual reason so far for the choice of
the gauge group $G$ as well as for the various representations
involved in the construction of SM. Thus under that light the
Standard Model appears as one of a plethora of possible quantum
field theories, and then needs to be minimally coupled to Einstein
gravity.

Our goal in this paper is to show that, in fact, the Standard Model
minimally coupled with Einstein gravity appears naturally as pure
gravity on a space $M\times F$ where the finite geometry $F$ is one
of the simplest and most natural finite noncommutative geometries of
the right dimension ($6$ modulo $8$) to solve the fermion doubling
problem.

Such a geometry is given by the following data:
\begin{itemize}
  \item A finite dimensional Hilbert space $\cH$
  \item An antilinear isometry $J$ of $\cH$ with $J^2=\epsilon$
  \item An involutive algebra $\cA$
  (over $\R$) acting in $\cH$, which fulfills the order zero condition
   \begin{equation}\label{order0first}
   [a,b^0]=0\qqq a,b\in \cA\,, \  b^0= J b^*J^{-1}\,.
   \end{equation}
  \item A $\Z/2$-grading $\gamma$ of $\cH$, such that
  $J\gamma=\epsilon''\gamma J$
  \item A self-adjoint operator $D$ in $\cH$ such that $JD=\epsilon' DJ$
\end{itemize}
In this paper we take the commutation relations \ie the values of
$(\epsilon,\epsilon',\epsilon'')\in \{\pm 1\}^3$ to be specific of
$K$-theoretic dimension $6$ modulo $8$ \ie
$(\epsilon,\epsilon',\epsilon'')=(1,1,-1)$. The reason for this
choice is that the product geometry $M\times F$ is then of
$K$-theoretic dimension $10$ modulo $8$ which allows one to use the
antisymmetric bilinear form $\langle J \xi,D\eta \rangle$ (for $\xi
,\eta \in \cH, \gamma \xi=\xi,\gamma\eta=\eta$) to define the
fermionic action, so that  the functional integral over fermions
delivers a Pfaffian rather than a determinant. In other words the
``raison d'être" for crossing by $F$ is to shift the $K$-theoretic
dimension from $4$ to $10$ (modulo $8$).

From the mathematical standpoint our road to $F$ is through the
following steps
\begin{enumerate}
  \item We classify the irreducible triplets $(\cA,\cH,J)$.
  \item We study the $\Z/2$-gradings $\gamma$ on $\cH$.
  \item We classify the subalgebras $\cA_F\subset \cA$ which allow for an
   operator $D$ that does not commute with the center of $\cA$ but fulfills the
   ``order one" condition:
   \begin{equation}\label{order1first}
[[D,a],b^0] = 0 \qquad \forall \, a,b \in \cA_F  \,.
\end{equation}
\end{enumerate}
The classification in the first step shows that the solutions fall
in two classes.

In the first case the solution is given by an integer $k$ and a real
form of the algebra $M_k(\C)$. The representation is given by the
action by left multiplication on $\cH=M_k(\C)$, and the isometry $J$
is given by $x\in M_k(\C)\mapsto J(x)= x^*$. There are three real
forms: unitary: $M_k(\C)$, orthogonal: $M_k(\R)$, symplectic:
$M_a(\H)$ where $\H$ is the skew field of quaternions, and $2a=k$.

In the second case the algebra is a real form of the sum
$M_k(\C)\oplus M_k(\C)$ of two copies of $M_k(\C)$ and while the
action is still given by left multiplication on $\cH=M_k(\C)\oplus
M_k(\C)$, the operator $J$ is given by $J(x,y)=(y^*,x^*)$.

The study (2) of the $\Z/2$-grading shows that the commutation
relation $J\gamma=-\gamma J$ excludes the first case. We are thus
left only with the second case and, after considering the grading we
are left with the symplectic--unitary algebra: $\cA=M_2(\H)\oplus
M_4(\C)$. At a more invariant level the Hilbert space is then of the
form $\cH=\Hom_\C(V,W)\oplus \Hom_\C(W,V)$ where $V$ is a
$4$-dimensional complex vector space, and $W$ a two dimensional
graded right vector space over $\H$. The left action of
$\cA=\End_\H(W)\oplus \End_\C(V)$ is by composition and its grading
as well as the grading of $\cH$ come from the grading of $W$.

Our main result then is that there exists up to isomorphism a unique
involutive subalgebra of maximal dimension $\cA_F$ of $\cA^{\rm
ev}$, the even part\footnote{One restricts to the even part to
obtain an ungraded algebra.} of the algebra $\cA$, which solves (3).
This involutive algebra $\cA_F $ is isomorphic to $\C\oplus \H
\oplus M_3(\C)$ and together with its representation in
$(\cH,J,\gamma)$  gives the noncommutative geometry $F$ of
\cite{mc2}.

We can then rely on the results of \cite{mc2}, which show that
(after the introduction of the multiplicity $3$ as the number of
generations) the spectral action applied to the inner fluctuations
on the product $M\times F$ delivers the Standard Model minimally
coupled to gravity. We refer to \cite{mc2} for the predictions which
follow using the spectral action at unification scale.

\section{The order zero condition and irreducible pairs $(\cA,J)$}

We start with a finite dimensional Hilbert space $\cH$  endowed with
an antiunitary operator $J$ such that $J^2= 1$. For any operator $x$
in $\cH$ we let,
\begin{equation}\label{xop}
x^0= J x^*J^{-1}\,.
\end{equation}
We look for involutive algebras $\cA$ of operators in $\cH$ such
that (\cf \eqref{order0first}),
\begin{equation}\label{orderzero}
[x,y^0]=0 \qqq x, y\in \cA\,.
\end{equation}
and that the following two conditions hold:
\begin{enumerate}
  \item The action of $\cA$ has a separating vector\footnote{\ie
  $\exists \xi\in \cH$ such that $\cA'\xi=\cH$ where $\cA'$ is the
  commutant of $\cA$.}
  \item The representation of $\cA$ and $J$ in $\cH$ is irreducible.
 \end{enumerate}
The role of the first condition is to abstract a natural property of
the action of an  algebra of (smooth) functions on the sections of a
vector bundle.

The meaning of the second condition is that one cannot find a
non-trivial projection $e\in \cL(\cH)$ which commutes with $\cA$ and
$J$.

\begin{lem} \label{center}  Assume  conditions \eqref{orderzero} and
(1), (2), then,

For any projection $e\neq 1$ in the center $Z(\cA)$ of $\cA$, one
has
\begin{equation}\label{noproj}
e JeJ^{-1}=0\,.
\end{equation}
For any projections $e_j$ in $Z(\cA)$ such that $e_1e_2=0$ one has
\begin{equation}\label{biproj}
e_1 Je_2J^{-1}+e_2 Je_1J^{-1}\in \{0,1\}\,.
\end{equation}
\end{lem}

\proof Let us show \eqref{noproj}. The projection $e JeJ^{-1}$
commutes with $\cA$ and $J$ since $JeJ^{-1}$ commutes with $\cA$ by
\eqref{orderzero}, and $J(e JeJ^{-1})= J( JeJ^{-1}e)=e J^{-1}
e=eJe=(e JeJ^{-1})J$.  Thus by irreducibility the projection $e
JeJ^{-1}$ is equal to $0$ or $1$ but the latter contradicts $e\neq
1$ since the range of $e JeJ^{-1}$ is contained in the range of $e$.

Let us show \eqref{biproj}. Since  $e_1e_2=0$ the sum $e_1
Je_2J^{-1}+e_2 Je_1J^{-1}$ is a projection and by the above argument
it commutes with $\cA$ and $J$. Thus by irreducibility it is equal
to $0$  or to $1$.\endproof

We let $\cA_\C$ be the complex linear space generated by $\cA$ in
the algebra $\cL(\cH)$ of all operators in $\cH$. It is an
involutive complex subalgebra of $\cL(\cH)$ and conditions
\eqref{orderzero}, (1) and (2) are still fulfilled.

\begin{lem} \label{dicho}  Assume  conditions \eqref{orderzero} and
(1), (2), then one of the following cases holds
\begin{itemize}
  \item The center $Z(\cA_\C)$ is reduced to $\C$.
  \item One has $Z(\cA_\C)=\C\oplus \C$ and $Je_1J^{-1}=e_2$ where
  $e_j\in Z(\cA_\C)$ are the minimal projections of $Z(\cA_\C)$.
\end{itemize}
\end{lem}

\proof Let us assume that the center $Z(\cA_\C)$ is not reduced to
$\C$. It then contains a partition of unity in minimal projections
$e_j$ with $\sum e_j=1$. By \eqref{noproj} we get
$$\sum_{i\neq j} e_i Je_jJ^{-1}=1\,.$$
The $e_i Je_jJ^{-1}$ are pairwise orthogonal projections, thus by
\eqref{biproj} there is a unique pair of indices $\{i,j\}=\{1,2\}$
such that
\begin{equation}\label{euntwo}
e_1 Je_2J^{-1}+e_2 Je_1J^{-1}=1\,,
\end{equation}
while the same expression vanishes for any other pair. For $i\notin
\{1,2\}$, one has $e_i=\sum e_i Je_kJ^{-1}=0$. It follows that all
other $e_i$ are zero and thus $Z(\cA_\C)=\C\oplus \C$. Moreover
since $e_1+e_2=1$, \eqref{euntwo} shows that $Je_2J^{-1}\geq e_1$
and $Je_1J^{-1}\geq e_2$ thus $Je_1J^{-1}=e_2$ and $Je_2J^{-1}=e_1$.
\endproof
\begin{rem}\label{metha1} Note that the above statements apply equally well in
case $J^2=\epsilon \in \{\pm 1\}$.
\end{rem}
Thus the classification of irreducible pairs splits in  the two
cases of Lemma \ref{dicho}.

\subsection{The case $Z(\cA_\C)=\C$}\hfill \medskip

We assume $Z(\cA_\C)=\C$. Then (\cf \cite{dix}) there exists
$k\in\N$ such that $\cA_\C=M_k(\C)$ as an involutive algebra over
$\C$. Moreover the algebra homomorphism
\begin{equation}\label{homo}
\cA_\C\otimes \cA_\C^0\to \cL(\cH)\,, \ \beta(x\otimes y)= xy^0 \qqq
x, y\in \cA_\C\,,
\end{equation}
is injective since $\cA_\C\otimes \cA_\C^0\sim M_{k^2}(\C)$ is a
simple algebra.

\begin{lem} \label{irred} The representation $\beta$ of $\cA_\C\otimes \cA_\C^0$
in $\cH$ of \eqref{homo} is irreducible.
\end{lem}

\proof Since $\cA_\C\otimes \cA_\C^0\sim M_{k^2}(\C)$, the
representation $\beta$ is a multiple of the unique representation
given by the left and right action of $\cA_\C=M_k(\C)$ on itself. We
need to show that the multiplicity $m$  is equal to $1$. We let $e$
be a minimal projection of $\cA_\C=M_k(\C)$, and let $E= e Je
J^{-1}$. By construction $E$ is a minimal projection of
$\cB=\cA_\C\otimes \cA_\C^0\sim M_{k^2}(\C)$ and thus its range has
dimension $m$. Moreover, by construction, $E$ commutes with $J$ so
that $J$ restricts to an antilinear isometric involution of square
$1$ on $E\cH$. Thus $E\cH$ is the complexification of a real Hilbert
space and $J$ the corresponding complex conjugation. Hence the
algebra of endomorphisms of $E\cH$ which commute with $J$ is
$M_m(\R)$ and\footnote{This is the only place where we use the
hypothesis $J^2=1$.}, if $\dim E\cH>1$, it contains a non-trivial
idempotent $F$. For any $\xi\in F\cH$, $\eta\in (E-F)\cH$ and $b\in
\cB$ one has
\begin{equation}\label{ortho}
\langle b\,\xi,\eta\rangle=0
\end{equation}
since, as $E$ is a minimal projection of $\cB$ one has $EbE=\lambda
E$ for some $\lambda\in \C$, and $\langle b\,\xi,\eta\rangle=\langle
EbE\,\xi,\eta\rangle=0$.  Thus $\cB F\cH$ is a non-trivial subspace
which is invariant under $\cB$ and $J$ since $J\cB J^{-1}=\cB$ and
$J$ commutes with $F$. This contradicts the irreducibility condition
(2).\endproof

\medskip
\begin{prop}\label{firstcase}
Let $\cH$ be a Hilbert space of dimension $n$. Then an irreducible
solution with $Z(\cA_\C)=\C$ exists iff $n=k^2$ is a square. It is
given by $\cA_\C=M_k(\C)$ acting by left multiplication on itself
and antilinear involution
\begin{equation}\label{invo}
 J(x)= x^* \qqq
x \in M_k(\C)\,. \end{equation}
\end{prop}

\proof We have $\cA_\C\otimes \cA_\C^0\sim M_{k^2}(\C)$ and by Lemma
\ref{irred} the representation $\beta$ in $\cH$ is irreducible. This
shows that $n=k^2$ is a square. The action of $\cA_\C\otimes
\cA_\C^0$ by left and right multiplication on $\cA_\C=M_k(\C)$
(endowed with the Hilbert-Schmidt norm) is a realization of the
unique irreducible representation of $\cA_\C\otimes \cA_\C^0\sim
M_{k^2}(\C)$. In that realization the canonical antiautomorphism
\begin{equation}\label{homo1}
\sigma(a\otimes b^0)=b\otimes a^0
\end{equation}
is implemented by the involution $J_0$, $J_0(x)=x^*$ of \eqref{invo}
\ie one has
$$
\sigma(x)=J_0 \,x^*J_0^{-1}\qqq x\in \cA_\C\otimes \cA_\C^0\,.
$$
Since the same property holds for the involution $J$ of the given
pair $(\cA,J)$ once transported using the unitary equivalence of the
representations, it follows that the ratio $J_0^{-1} J$ commutes
with $\cB$ and hence is a scalar $\lambda\in \C$ of modulus one by
irreducibility of $\beta$. Adjusting the unitary equivalence by a
square root $\mu$ of $\lambda$ (using $\mu J \mu^{-1}=\mu^2 J$) one
can assume that $J=J_0$ which gives the desired uniqueness.
\endproof

\medskip

This determines $\cA_\C$ and its representation in $(\cH,J)$ and it
remains to list the various possibilities for $\cA$. Now $\cA$ is an
involutive subalgebra of $M_k(\C)$ such that $\cA+i\cA=M_k(\C)$. The
center $Z(\cA)$ is contained in $Z(M_k(\C))=\C$. If $Z(\cA)=\C$ then
$i\in \cA$ and $\cA=M_k(\C)$. Otherwise one has $Z(\cA)=\R$, $\cA$
is a central simple algebra over $\R$ (the simplicity follows from
that of $\cA+i\cA=M_k(\C)$) and $\cA\cap i\cA=\{0\}$ (since this is
an ideal in $\cA$). Thus $\cA$ is the fixed point algebra of the
antilinear automorphism $\alpha$ of $M_k(\C)$ commuting with the
$*$-operation, given by $\alpha(a+i b)=a-ib$ for $a,b\in \cA$. There
exists (comparing $\alpha$ with complex conjugation) an antilinear
isometry $I$ of $\C^k$ such that $\alpha(x)=IxI^{-1}$ for all $x\in
M_k(\C)$. One has $\alpha^2=1$ and thus $I^2\in \{\pm 1\}$ (it is a
scalar $\lambda\in \C$ of modulus one and commutes with $I$). Thus
$\cA$ is the commutant of $I$ and the only two cases are $I^2=1$
which gives matrices $M_k(\R)$ over $\R$ and $I^2=-1$. In the latter
case the action of $I$ turns $\C^k$ into a right vector space over
the quaternions $\H$ and  $k=2a$, $\cA=M_a(\H)$ is the algebra of
endomorphisms of this vector space over $\H$.  We can thus summarize
the three possibilities
\begin{itemize}
  \item $\cA=M_k(\C)$ (unitary case)
  \item $\cA=M_k(\R)$  (orthogonal case)
  \item $\cA=M_a(\H)$, for even $k=2a$, (symplectic case)
\end{itemize}
while the representation is by left multiplication on $M_k(\C)$ and
the antilinear involution $J$ is given by \eqref{invo}.

\begin{rem}\label{metha2} Note that in the case $J^2=-1$
the possibility of multiplicity $m=2$ arises and the dimension of
$\cH$ is $2k^2$ in that case.
\end{rem}

We shall prove below in Lemma \ref{gradecase1} that the above case
$Z(\cA_\C)=\C$ is incompatible with the commutation relation
$J\gamma=-\gamma J$ for the grading and hence with the
$K$-theoretic dimension $6$. Thus we now concentrate on the second
possibility: $Z(\cA_\C)=\C\oplus \C$.

\medskip

\subsection{The case $Z(\cA_\C)=\C\oplus \C$}\hfill \medskip

We assume $Z(\cA_\C)=\C\oplus \C$. Then there exists $k_j\in\N$ such
that $\cA_\C=M_{k_1}(\C)\oplus M_{k_2}(\C)$ as an involutive algebra
over $\C$. We let $e_j$ be the minimal projections $e_j\in
Z(\cA_\C)$ with $e_j$ corresponding to the component $M_{k_j}(\C)$.

\smallskip
\begin{lem} \label{irredbis}
\begin{enumerate}
\item  The representation $\beta$ of $\cA_\C\otimes \cA_\C^0$ in $\cH$
of  \eqref{homo} is the direct sum of two irreducible
representations in the decomposition
\begin{equation}\label{decofh}
\cH=e_1\cH\oplus e_2\cH=\cH_1\oplus \cH_2\,,\
\beta=\beta_1\oplus\beta_2.
\end{equation}
\item The representation $\beta_1$ (resp $\beta_2$) is the only irreducible
representation of the reduced algebra of $\cB$ by $e_1\otimes e_2^0$
(resp. $e_2\otimes e_1^0$).
\item The dimension of $\cH_j$ is equal to $k_1k_2$.
\end{enumerate}
\end{lem}

\proof 1) Let $\cH_j=e_j\cH$. Since $e_j\in Z(\cA)$  the action of
$\cA$ in $\cH$ is diagonal in the decomposition \eqref{decofh}. By
Lemma \ref{dicho} one has $Je_jJ^{-1}=e_k$, $k\neq j$, thus the
action of $\cA^0$ is also diagonal in the decomposition
\eqref{decofh}. Thus the representation $\beta$ decomposes as a
direct sum $\beta=\beta_1\oplus\beta_2$. Moreover by Lemma
\ref{dicho} the antilinear involution $J$ interchanges the $\cH_j$.
Let $F_1$ be an invariant subspace for the action $\beta_1$ of $\cB$
in $\cH_1$. Then $F_1\oplus JF_1\subset \cH$ is invariant under both
$\cB$ and $J$ and thus equal to $\cH$ by irreducibility which
implies $F_1=\cH_1$.

2) This follows since in each case one gets an irreducible
representation of the reduction of $\cB$ by the projections
$e_i\otimes e_j^0$, $i\neq j$, which as an involutive algebra is
isomorphic to $M_{k_i}(\C)\otimes M_{k_j}(\C)\sim M_{k_1k_2}(\C) $.

3) This follows from 2).
\endproof

\medskip
\begin{prop} \label{case2}
Let $\cH$ be a Hilbert space of dimension $n$. Then an irreducible
solution with $Z(\cA_\C)=\C\oplus \C$ exists iff $n=2k^2$ is twice a
square. It is given by $\cA_\C=M_k(\C)\oplus M_k(\C)$ acting by left
multiplication on itself and antilinear involution
\begin{equation}\label{involbis}
 J(x,y)=(y^*,x^*) \qqq
x, y\in M_k(\C)\,.
 \end{equation}
\end{prop}

\proof Let us first show that $k_1=k_2$. The dimension of $\cA_\C$
is $k_1^2+k_2^2$. The dimension of $\cH$ is $2 k_1k_2$ by Lemma
\ref{irredbis}. The separating condition  implies $\dim \cA_\C\leq
\dim \cH$ because of the injectivity of the map $a\in \cA_\C\mapsto
a\xi\in \cH$ for $\xi$ such that $\cA'\xi=\cH$. This gives
$$
k_1^2+k_2^2\leq 2 k_1k_2
$$
which is possible only if $k_1=k_2$.  In particular $n=2k^2$ is
twice a square. We have shown that $\cA_\C=M_k(\C)\oplus M_k(\C)$
and moreover by Lemma \ref{irredbis} the representation $\beta$ is
the direct sum of the irreducible representations of the reduced
algebras of $\cB$ by the projections $e_1\otimes e_2^0$ and
$e_2\otimes e_1^0$. Thus we can assume that the representation
$\beta$ of $\cB$ is the same as in the model of Proposition
\ref{case2}. It remains to determine the antilinear isometry $J$. We
let $J_0(x,y)=(y^*,x^*) \qqq x, y\in M_k(\C)$ as in \eqref{involbis}
and compare the antilinear isometry $J$ of the given pair with
$J_0$. By the argument of the proof of Proposition \ref{firstcase},
we get that the ratio $J_0^{-1} J$ commutes with $\cB$ and hence is
a diagonal matrix of scalars $\left(
                                \begin{array}{cc}
                                  \lambda_1 & 0 \\
                                  0 & \lambda_2 \\
                                \end{array}
                              \right)$
                              in the decomposition $\cH=\cH_1\oplus
                              \cH_2$.
The condition $J^2=1$ shows that $\lambda_1=\lambda_2$. Thus
$J=\lambda J_0$ and the argument of the proof of Proposition
\ref{firstcase} applies to give the required uniqueness.
\endproof

\medskip

\begin{rem} \label{basefree}  {\rm  One can describe
 the above solutions  (\ie  the algebra $\cA_\C$ and its representation in
$\cH, J$) in a more intrinsic manner as follows. We let $V$  and $W$
be $k$-dimensional complex Hilbert spaces. Then
\begin{equation}\label{algebra0}
\cA_\C=\End_\C(W)\oplus \End_\C(V) \,.
\end{equation}
We let $\cH$ be the bimodule over  $\cA_\C$ given by
\begin{equation}\label{module2}
\cH=\cE\,\oplus \,\cE^*\,,\ \ J(\xi,\eta)=(\eta^*,\xi^*)
\end{equation}
where,
\begin{equation}\label{module1}
\cE=\Hom_\C (V, W)\,,  \ \ \cE^*=\Hom_\C (W, V)
\end{equation}
and the algebra acts on the left by composition:
\begin{equation}\label{module3}
(w,v)(g,h)=( w\circ g,v\circ h) \qqq (w,v)\in \cA_\C\,,\ (g,h)\in
\cE\,\oplus \,\cE^*\,.
\end{equation}
The various real forms can then be described using additional
antilinear isometries of $V$ and $W$. }\end{rem}

\medskip

\section{$\Z/2$-grading}

In the set-up of spectral triples one assumes that in the even case
the Hilbert space is $\Z/2$-graded \ie endowed with a grading
operator $\gamma$, $\gamma^2=1$, $\gamma=\gamma^*$. This grading
should be compatible with a $\Z/2$-grading of the algebra $\cA$
which amounts to asking that $\gamma \cA \gamma^{-1}=\cA$. One then
has $[\gamma,a]=0$ for any $a\in \cA^{\rm ev}$ the even part of
$\cA$.

\begin{lem} \label{gradecase1}
In the case $Z(\cA_\C)=\C$ of Proposition
\ref{firstcase}, let $\gamma$ be a $\Z/2$-grading of $\cH$ such that
$\gamma \cA \gamma^{-1}=\cA$ and $J\gamma=\epsilon'' \gamma J$ for
$\epsilon''=\pm 1$. Then $\epsilon''=1$.
\end{lem}

\proof We can assume that $(\cA_\C,\cH,J)$ are as in Proposition
\ref{firstcase}. Let $\delta \in \Aut(\cA_\C)$ be the automorphism
given by $ \delta(a)=\gamma a \gamma^{-1}$, $\forall a \in\cA_\C\,.
$ One has $\delta^2=1$. Similarly one gets an automorphism
$\delta^0$ of $\cA_\C^0$ such that $ \delta^0(b^0)=\gamma b^0
\gamma^{-1}$ since $\gamma \cA_\C^0 \gamma^{-1}=\cA_\C^0$ using the
relation $J\gamma=\epsilon'' \gamma J$. Then $\delta\otimes
\delta^0$ defines an automorphism of $\cA_\C\otimes \cA_\C^0$ such
that
$$
\gamma \beta(x) \gamma^{-1}=\beta(\delta\otimes \delta^0(x))\qqq
x\in \cB=\cA_\C\otimes \cA_\C^0\,.
$$
Thus $\gamma$ implements the tensor product of two automorphisms of
$M_k(\C)$. These automorphisms are inner and it follows that there
are unitary matrices $u,v\in M_k(\C)$ such that $ \gamma(a) = u a
v^*$, $\forall a \in\cA_\C\,. $ One then has
$$J\gamma J^{-1}(a)=(u a^* v^*)^*=vau^*\qqq a \in M_k(\C)\,.$$ Thus
the equality $J\gamma J^{-1}=-\gamma$ means that
$$
vau^*=-u a v^*\qqq a \in M_k(\C)\,,
$$
\ie that $u^*v=z$ fulfills $zaz=-a$ for all $a$. Thus $z^2=-1$ and
$za=az$ for all $a$ so that $z=\eta i$ for some $\eta\in\{\pm 1\}$.
We thus get $v=\eta i u$. Then $\gamma(a)=-\eta i uau^{*}$ and
$\gamma^{2}(a)=-u^2au^{-2}$. But since $\gamma^2=1$ one gets that
$a=-u^2au^{-2}$ for all $a\in M_k(\C)$ which is a contradiction for
$a=1$.
\endproof

Thus Lemma \ref{gradecase1} shows that we cannot obtain the required
commutation relation $J\gamma=-\gamma J$ in the case $Z(\cA_\C)=\C$.

\medskip
At this point we are at a cross-road. We know that we are in the
case $Z(\cA_\C)=\C\oplus \C$ but we must choose the integer $k$ and
the real form $\cA$ of $\cA_\C=M_k(\C)\oplus M_k(\C)$.

We make the {\em hypothesis} that  both the grading and the real
form come by assuming that the vector space $W$ of Remark
\ref{basefree} is  a right vector space over $\H$ and is
non-trivially $\Z/2$-graded. The right action of quaternions amounts
to giving an antilinear isometry $I$ of $W$ with $I^2=-1$ (\cf
\cite{mehta} Chapter 3). Since $W$ is a non-trivially $\Z/2$-graded
vector space over $\H$ its dimension must  be at least $2$ (and
hence $4$ when viewed as a complex vector space). We choose the
simplest case \ie $W$ is a two-dimensional space over $\H$, and
there is no ambiguity since all non-trivial $\Z/2$-gradings are
equivalent.
 A conceptual description of
 the algebra $\cA$ and its representation in
$\cH$ is then obtained from Remark \ref{basefree}. We let $V$ be a
$4$-dimensional complex vector space. Our algebra is
\begin{equation}\label{algebra}
\cA=\End_\H(W)\oplus\End_\C(V) \sim M_2(\H)\oplus  M_4(\C)  \,.
\end{equation}
 It follows from the grading of $W$ that the algebra \eqref{algebra} is also $\Z/2$-graded,
with non-trivial grading only on the $M_2(\H)$-component. We still
denote by $\gamma$ the gradings of $\cE=\Hom_\C (V, W)$ and
$\cE^*=\Hom_\C (W, V)$ given by composition with the grading of $W$.
We then have, with the notations of \eqref{module2},

\begin{prop} \label{finalform} There exists up to equivalence a unique $\Z/2$-grading
of $\cH$ compatible with the graded representation of $\cA$ and such
that:
\begin{equation}\label{dim6}
J\,\gamma=-\gamma \, J
\end{equation}
It is given by
\begin{equation}\label{modulegrad}
\cH=\cE\,\ominus \,\cE^*\,,\ \ \gamma(\xi,\eta)=(\gamma\xi,
-\gamma\eta)
\end{equation}
\end{prop}
\proof By construction the grading \eqref{modulegrad} is a solution.
Given two gradings $\gamma_j$ fulfilling the required conditions one
gets that their ratio $\gamma_1\gamma_2$ commutes with $\cA$ (since
both define the grading of $\cA$ by conjugation) and with $J$. Thus
by irreducibility one gets that $\gamma_1\gamma_2\in \pm 1$.
Changing $\gamma$ to $-\gamma$ amounts to changing the grading of
$W$ to its opposite, but up to isomorphism this gives the same
result.
\endproof

\medskip
\begin{rem}\label{k=4} {\rm The space $\cE=\Hom_\C (V, W)$
is related to the classification of instantons (\cf Equation (1.1)
Chapter III of \cite{atiyah}).}
\end{rem}

\medskip
\section{The subalgebra and the order one condition}

We take $(\cA,\cH,J,\gamma)$ from the above discussion, \ie
\eqref{algebra}  and Proposition \ref{finalform}.

The center of our algebra $Z(\cA)$ is non-trivial and in that way
the corresponding space is not connected. We look for ``Dirac
operators" $D$ which connect non-trivially the two pieces (we call
them ``off-diagonal") \ie operators such that:
\begin{equation}\label{offdiag}
[D,Z(\cA)]\neq \{ 0\}  .
\end{equation}
The main requirement on such operators is the order one condition
\eqref{order1first}. We now look for subalgebras $\cA_F \subset
\cA^{\rm ev}$, the even part of $\cA$, for which this order
condition ($\forall \, a,b \in \cA_F$)  allows for operators which
fulfill \eqref{offdiag}. We can now state our main result which
recovers in a more conceptual manner the main ``input" of
\cite{mc2}.
\medskip

\begin{thm}\label{nameless} Up to an automorphism of $\cA^{\rm ev}$,
there exists a unique involutive subalgebra $\cA_F \subset \cA^{\rm
ev}$ of maximal dimension admitting off-diagonal Dirac operators. It
is given by
\begin{equation}\label{subalgF}
\cA_F =\{ (\lambda \oplus q,\lambda\oplus m)\;|\;\lambda\in \C\,,\;
q\in \H \,,\; m\in M_3(\C)\}\subset \H\oplus \H\oplus M_4(\C)\,,
\end{equation}
using  a field morphism $\C\to \H$. The involutive  algebra $\cA_F $
is isomorphic to $\C\oplus \H \oplus M_3(\C)$ and together with its
representation in $(\cH,J,\gamma)$ it gives the noncommutative
geometry $F$ of \cite{mc2}.
\end{thm}

We now give the  argument  (which is similar to that of  \cite{mc2}
Proposition 2.11).
 Let us consider the
decomposition of \eqref{decofh},
$$
\cH=e_1\cH\oplus e_2\cH=\cH_1\oplus \cH_2\,.
$$
We consider an involutive subalgebra $\cA_F \subset \cA^{\rm ev}$
and let $\pi_j$ be the restriction to $\cA_F$ of the representation
of $\cA$ in $\cH_j$. We have (\cf \cite{mc2} Lemma 2.12),

\begin{lem} \label{lemoffdiag}
If the representations  $\pi_j$  are disjoint, then there is no off
diagonal Dirac operator for $\cA_F$.
\end{lem}

\proof By construction the projections $e_j$ are the minimal
projections in $Z(\cA)$ and since $Je_1J^{-1}=e_2$, they are also
the minimal projections in $Z(\cA^0)$. Moreover one has
$$
\pi_j(a)=a e_j=e_j a=e_j a e_j \qqq a \in \cA_F\,.
$$
Let us assume that the representations  $\pi_j$  are disjoint. For
any operator $T$ in $\cH$, one has
\begin{equation}\label{disjres}
[T,a]=0 \qqq a \in \cA_F \Rightarrow [e_1Te_2,a]=0 \qqq a \in \cA_F
\Rightarrow e_1Te_2=0\,,
\end{equation}
since any intertwining operator such as $e_1Te_2$ must vanish as the
two representations are disjoint. The same conclusion applies to
$e_2Te_1$. Similarly one gets, after conjugating by $J$,
\begin{equation}\label{disjresbis}
[T,a^0]=0 \qqq a \in \cA_F \Rightarrow e_1Te_2=e_2Te_1=0\,.
\end{equation}
(one has $[JTJ^{-1},a]=0$   for all $a\in \cA$ hence
$e_2JTJ^{-1}e_1=0$ and $e_1 T e_2=0$).
Now let the operator $D$
satisfy the order one condition
\begin{equation}\label{order1bis}
[[D,a],b^0] = 0 \qquad \forall \, a,b \in \cA_F  \, .
\end{equation}
It follows from \eqref{disjresbis} that $e_1[D,a]e_2=0$ for all
$a\in \cA_F$. Since the $e_j$ commute with $a$ this gives
$$
[e_1De_2,a]=0 \qqq a \in \cA_F  \, .
$$
By \eqref{disjres} we thus get $e_1De_2=0$ and there is no off
diagonal Dirac operator for $\cA_F$.\endproof

For any operator $T\;:\;\cH_1\to \cH_2$ we let
\begin{equation}\label{comrel}
\cA(T)=\{ b\in \cA^{\rm ev}\;|\;
\pi_2(b)T=T\pi_1(b)\,,\;\pi_2(b^*)T=T\pi_1(b^*)\} .
\end{equation}
It is by construction an involutive unital subalgebra of $\cA^{\rm
ev}$.

We now complete the proof of Theorem \ref{nameless}. We let $\cA_F
\subset \cA^{\rm ev}$ be an involutive subalgebra with an off
diagonal Dirac operator. Then by Lemma \ref{lemoffdiag},  the
representations $\pi_j$ are not disjoint and thus there exists a
non-zero operator $T\;:\;\cH_1\to \cH_2$ such that $\cA_F \subset
\cA(T)$. If we replace $T\to c_2Tc_1$ where $c_j$ belongs to the
commutant of $\cA^{\rm ev}$, we get
$$
\cA(T) \subset \cA(c_2Tc_1)\,,
$$
since the $c_j$ commute with the $\pi_j(b)$. This allows one to
assume that the support of $T$ is contained in an irreducible
subspace of the restriction of the action of $\cA^{\rm ev}$ on
$\cH_1$ and that the range of $T$ is contained in an irreducible
subspace of the restriction of the action of $\cA^{\rm ev}$ on
$\cH_2$. We can thus  assume that $\pi_1$ is the irreducible
representation of one of the two copies of $\H$ in $\C^2$, while
$\pi_2$ is the irreducible representation of $M_4(\C)$ in $\C^4$. We
can remove the other copy of $\H$ and replace $\cA^{\rm ev}=\H\oplus
\H \oplus M_4(\C)$ by its projection $\cC=\H \oplus M_4(\C)$. The
operator $T$ is a non-zero operator $T\;:\;\C^2\to \C^4$ and with
\begin{equation}\label{comrelbis}
\cC(T)=\{ b\in \cC\;|\;
\pi_2(b)T=T\pi_1(b)\,,\;\pi_2(b^*)T=T\pi_1(b^*)\} .
\end{equation}
we have that $\cA(T)=\{(q,y)\,|\,q\in \H\,,\ y\in \cC(T)\}$. In
particular $\dim \cA(T)= 4 + \dim \cC(T)$. Let us first assume that
the rank of $T$ is equal to $2$. The range $E$ of $T$ is a two
dimensional subspace of $\C^4$ and by \eqref{comrelbis} it is
invariant under the action of $b\in \cC$ as well as its orthogonal
complement. This shows that in that case
$$
\cC(T)\subset \H\oplus M_2(\C)\oplus M_2(\C)\,,
$$
and moreover the relation \eqref{comrelbis} shows that the component
of $\pi_2(b)$ in the copy of $M_2(\C)$ corresponding to $E$ is
determined by the quaternion component $\pi_1(b)$. Thus we get
$$
\cC(T)\subset \H\oplus M_2(\C)\,.
$$
In particular the dimension fulfills
$$
\dim_\R \cC(T)\leq 4+8=12\,.
$$
Let us now consider the other possibility, namely that the rank of
$T$ is equal to $1$. The range $E$ of $T$ is a one dimensional
subspace of $\C^4$ and by \eqref{comrelbis} it is invariant under
the action of $b\in \cC$ as well as its orthogonal complement. The
support $S\subset \C^2$ of $T$ is a one dimensional subspace and
since both the unitary group $\SU(2)$ of $\H$ and ${\rm U}(4)$ of
$M_4(\C)$ act transitively on the one dimensional subspaces (of
$\C^2$ and $\C^4$) we are reduced to the case
$$
S=\{(a,0)\in \C^2\}\,,\ \ E=\{(a,0,0,0)\in \C^4\}\,,\ \
T(a,b)=(a,0,0,0)\qqq a,b\in \C\,.
$$
One then obtains, for the natural embedding $$\C\subset \H\,,\
   \lambda \mapsto \left(
          \begin{array}{cc}
            \lambda & 0 \\
            0 & \bar\lambda \\
          \end{array}
        \right)
$$ that,
$$
\cC(T)=\{(\lambda,\lambda \oplus m)\in \H\oplus M_4(\C)\,|\, \lambda
\in \C\,, \ m\in M_3(\C)\} \,.
$$
Thus the dimension fulfills
$$
\dim_\R \cC(T)= 2+18=20\,.
$$
Thus we see that this gives the  solution  with maximal dimension,
and it is unique up to an automorphism of $\cA^{\rm ev}$. \endproof

\medskip
We can now combine the above discussion with the result of
\cite{mc2} Theorem 4.3 and get,

\begin{thm}\label{maintheorem}
Let $M$ be a Riemannian spin $4$-manifold and $F$ the finite
noncommutative geometry of $K$-theoretic dimension $6$ described above, but
with multiplicity\footnote{\ie we just take three copies of $\cH$}
$3$. Let $M\times F$ be endowed with the product metric.
\begin{enumerate}
\item The unimodular subgroup of the unitary group acting by the adjoint
representation ${\rm Ad}(u)$ in $\cH$ is the group of gauge
transformations of SM.
\item The unimodular inner fluctuations of the metric give the gauge
bosons of SM.
\item The full standard model (with neutrino mixing and seesaw
mechanism) minimally coupled to Einstein gravity is given in
Euclidean form by the action functional
\begin{equation}\label{functspec}
S=\,\Tr(f(D_A/\Lambda))+\frac
12\,\langle\,J\,\tilde\xi,D_A\,\tilde\xi\rangle\,,\quad\tilde\xi\in
\cH^+_{cl} ,
\end{equation}
where $D_A$ is the Dirac operator with the unimodular inner
fluctuations.
\end{enumerate}
\end{thm}

We refer to \cite{mc2} for the notations and for the predictions.

\begin{rem}\label{unimodular}{\rm
The ``unimodularity" condition
  imposed in Theorem \ref{maintheorem} (\cf \cite{mc2}) on our gauge transformations
  can now be viewed as the restriction to $\cA_F$ of the condition
  giving the group of
  inner automorphisms of $\cA$. Indeed
this group is described as the  unimodular unitary group:
  \begin{equation}\label{unimod}
\Int(\cA)\sim\SU(\cA)=\{ u\in \cA\,| uu^*=u^*u=1,\,\det(u)=1\}\,.
  \end{equation}
  This applies also to the product geometry by the manifold $M$.}
  \end{rem}

\bigskip
\section{Conclusion}

The fermion doubling problem requires (\cf \cite{Barrett},
\cite{CoSMneu}) crossing the ordinary $4$-dimensional continuum by a
space of $K$-theoretic dimension $6$. We have shown in this paper that the
classification of the finite noncommutative geometries of
$K$-theoretic dimension $6$ singles out the algebras which are real forms of
$M_k(\C)\oplus M_k(\C)$ acting in the Hilbert space of dimension
$2k^2$ by left multiplication, together with a specific antilinear
isometry. This predicts the number of fermions per generation to be
a square and under our {\em hypothesis} about the role of
quaternions the simplest case is with $k=4$ and gives the
noncommutative geometry of the standard model in all its details,
including the representations of fermions and bosons and the
hypercharges.

While we have been able to find a short path to the Standard Model
coupled to gravity from simple geometric principles  using
noncommutative geometry and the spectral action, there are still a
few forks along the way where the choice we made would require a
better justification.  The list is as follows:

\begin{description}
  \item[Why $\H$]  The field $\H$ of quaternions plays an important
  role in our construction, since we assumed that  both the grading and the real
form come from $W$ being quaternionic. This is begging for a better
  understanding. The role of quaternions in the classification of
  instantons (\cite{atiyah}) is one possible starting point as well
  as the role of discrete symmetries (\cf \cite{mehta}).
    \item[Three generations] We took the number $N=3$ of generations
  as an input which gave the multiplicity $3$. From the physics
  standpoint, violation of $CP$ is a reason for $N\geq 3$ but it
  remains to find a convincing mathematical counterpart.
    \item[Massless photon] In the classification (\cite{mc2}) of the operators $D$ for
    the finite geometry $F$, we impose that $D$ commutes with the
  subalgebra   $\{(\lambda,\lambda,0);\lambda\in \C\}$. While the physics meaning of
  this condition is clear since it amounts to the masslessness of
  the photon, a conceptual mathematical reason for only considering
  metrics fulfilling this requirement is out of sight at the moment.
\end{description}
Our approach delivers the unique representation for the fermions, a
property which is only shared with the ${\rm SO}(10)$ grand-unified
theory.   One of the main advantages of our approach with respect to
unified theories is that  the reduction to the Standard Model group
$G={\rm U}(1)\times \SU(2)\times \SU(3)$ is not due to a plethora of
added scalar Higgs fields, but is naturally imposed by the order one
condition.

The spectral action of the standard model comes out almost uniquely,
predicting the number of fermions, their representations, the gauge
group and their quantum numbers as well as the Higgs mechanism, with
very little input. This manages to combine the advantages of
Kaluza-Klein unification (we are dealing with pure gravity on a
space of $K$-theoretic dimension ten) with those of
grand-unification such as ${\rm SO}(10)$ (including the unification
of coupling constants) without introducing unobserved fields and an
infinite tower of states.

\medskip

{\bf Acknowledgments.--} The research of A. H. C. is supported in
part by the National Science Foundation under Grant No. Phys-0601213
and by a fellowship from the Arab Fund for Economic and Social
Development.

\end{document}